\documentclass[aps,prb,amsmath,amssymb,twocolumn]{revtex4-2}
\usepackage{graphicx}
\usepackage{bm}
\usepackage{mathptmx}
\usepackage{epstopdf}
\usepackage{booktabs}
\usepackage{multirow}
\usepackage{hhline}
\usepackage{subfigure}
\usepackage{tabularx}
\usepackage{float}
\usepackage{xcolor}
\begin{document}
\newcommand{\ms}{m_*}
\newcommand{\bor}{{\bf r}}
\newcommand{\mm}{{\bf m}}
\newcommand{\rdot}{\dot{\bf r}}
\newcommand{\rddot}{\dot\dot{\bf r}}
\newcommand{\es}{\bf{E^s_\sigma}}

\title{Multi-site mixing and entropy stabilization of CsPbI$_{3}$ with potential application in photovoltaics}
\author{Namitha Anna Koshi$^{1}$}
\author{Krishnamohan Thekkepat$^{2}$}
\author{Doh-Kwon Lee$^{3}$}
\author{Seung-Cheol Lee$^{2}$}
\email{leesc@kist.re.kr}
\author{Satadeep Bhattacharjee$^{1}$}
\email{s.bhattacharjee@ikst.res.in}
\affiliation{$^{1}$Indo-Korea Science and Technology Center (IKST), Bengaluru 560064, India}
\affiliation{$^{2}$Electronic Materials Research Center, Korea Institute of Science and Technology (KIST), Seoul 02792, Republic of Korea}
\affiliation{$^{3}$Advanced Photovoltaics Research Center, Korea Institute of Science and Technology (KIST), Seoul 02792, Republic of Korea}

\begin{abstract} 
Metal halide perovskite solar cells have achieved dramatic improvements in their power conversion efficiency in the recent past. Since compositional engineering plays an important role in optimizing material properties, we investigate the effect of alloying at Cs and Pb sites on the energetics and electronic structure of CsPbI$_{3}$ using cluster expansion method in combination with first-principles calculations. For Ge-mixing at Pb-site, the $\alpha$ and $\beta$-phases are considered with emphasis on the electronic structure, transition probability, absorption coefficient, efficiency, and carrier mobility of higher-symmetry configurations. CsPb$_{0.50}$Ge$_{0.50}$I$_{3}$ (Cs$_{2}$PbGeI$_{6}$) which takes up a double perovskite (elpasolite) structure has a direct band gap with no parity-forbidden transitions. 
Further, we utilize the alloy entropic effect to improve the material stability and optoelectronic properties of CsPbI$_{3}$ by multi-element mixing. For the proposed mixed compositions, the Fr{\"o}hlich electron-phonon coupling constant is determined. Scattering rates and electron mobility are obtained from first-principles inputs. These lower Pb-content inorganic perovskites offer great promise as efficient solar cell materials for photovoltaic applications.
\end{abstract}


\maketitle

\section{Introduction}
The need for sustainable and clean sources of energy has shifted the focus of the scientific community to solar cells based on metal halide perovskites, as their thin films prepared by solution processing have the advantage of reduced production costs \cite{xiang2021review}. Perovskite solar cells (PSCs) comprising of hybrid organic-inorganic and all-inorganic perovskites are emerging as competitive rivals for silicon solar cells, which dominate the photovoltaic (PV) market. These halide perovskites have the generic formula ABX$_{3}$, with the A-site occupied by organic cations or cesium (Cs), the B-site by Pb or Sn, and the X-site by the halogen ion. The most celebrated hybrid organic-inorganic perovskites are MAPbI$_{3}$ and FAPbI$_{3}$. The ever-improving power conversion efficiency (PCE) of these hybrid perovskites has garnered significant attention in recent years. Though they have great potential, there are serious challenges like cell instability against moisture and temperature and the presence of toxic lead (Pb), that hamper their real world deployment \cite{nagabhushana2016direct, aristidou2015role, babayigit2016toxicity, savory2016can}. All-inorganic lead halide perovskites have improved chemical stability than hybrid ones \cite{steele2019thermal, kulbak2016cesium, eperon2015inorganic}. Among all-inorganic perovskites (i.e. CsPbX$_{3}$), the black cubic phase of CsPbI$_{3}$ (called $\alpha$) has a solar-friendly gap of $\sim$1.7 eV with strong absorption in the UV-visible range of the solar spectrum. However, $\alpha$-CsPbI$_{3}$ is thermodynamically unstable at room temperature, and synergistic efforts are made to stabilize it. Among the many strategies that have been adopted include the partial replacement of I$^{-}$ with smaller sized Br$^{-}$ or Cl$^{-}$. However, this anion substitution has been found to be detrimental for solar cell efficiency as the band gap increases with Br/Cl substitution \cite{hoffman2016transformation, protesescu2015nanocrystals}. For replacing monovalent Cs$^{+}$, elements with a larger ionic radius are not common. Although organic cations have an effective ionic radius larger than that of cesium, they suffer from their volatile nature. 

Another route for designing stable and low Pb-content halide perovskite is compositional engineering of CsPbI$_{3}$ via B-site substitution. Among the divalent cations, Sn$^{2+}$, Ge$^{2+}$, and Si$^{2+}$ belonging to the same group as Pb, are more favorable for B-site substitution. Both Sn$^{2+}$ and Ge$^{2+}$ exhibit lone pair s orbitals similar to Pb-based perovskites \cite{meng2017parity}. Besides that, Sn-based halide perovskite (CsSnI$_{3}$) possess a lower band gap than CsGeI$_{3}$, which can be attributed to the Sn 5s orbital lying at higher energy than Ge 4s orbital. Since Ge 4s level sits closer to Pb 6s level, Pb-Ge mixing at B-site could result in structures with a band gap higher than the Pb-Sn mixed ones. The Goldschmidt tolerance factor ($t$) is one of the empirical parameters used to assess the structural stability of perovskites. The $t$ of CsGeI$_{3}$ is 0.93 compared to 0.81 of CsPbI$_{3}$. Hence, mixing Pb and Ge results in structures with higher $t$ than CsPbI$_{3}$. One major drawback of Sn and Ge-based perovskite solar cells as reported in previous studies, is the stability issue of 2+ oxidation state of Sn and Ge, which gets oxidized to 4+ state when exposed to air \cite{noel2014lead, hoshi2016improved, marshall2016enhanced, leijtens2017mechanism, zhao2017design}. Other divalent cations (such as Mg, Ca, Sr, Zn, Cd) with no lone pair s orbital results in indirect band gap semiconductors, which are not desirable for thin-film solar cell applications \cite{meng2017parity}. In addition, some divalent cations have the potential to form defect states in the resultant structure, which can act as traps or recombination centers \cite{noel2014lead, hao2014lead}, thereby affecting the PV performance. The complete replacement of Pb could affect the optoelectronic properties, therefore tuning the composition of the mixed compound is very essential \cite{park2015bismuth, pal2017colloidal}.

Entropy stabilization is a promising strategy in halide perovskites with applications in photovoltaics and thermoelectrics \cite{muzzillo2024high, jiang2021high}. In this approach, the large configurational entropy of multi-element alloys is used to overcome the mixing enthalpy barrier and boost the thermodynamical stability at finite temperatures \cite{wang2022entropy, deng2020semiconducting, muzzillo2024high}. In halide perovskite nanocrystals, entropy stabilization is demonstrated in CH$_{3}$NH$_{3}$PbBr$_{3}$ by substituting Mg, Zn and Cd at Pb-site with the preservation of narrow band emission \cite{solari2022stabilization}. A recent report on high entropy alloying in halide perovskites projected that mixing Br, Cl, and I on the X site, Cs and Rb on the A site and Ge, Pb, and Sn on the B site are all promising \cite{muzzillo2024high}. The greatest entropy stabilization comes with halide mixing, but it has the highest unit-cell volume coefficient of variation. These works inspired us to investigate A and B-site mixing in CsPbI$_{3}$ and explore the optoelectronic behavior of the resultant compositions. In this paper, we started by mixing Ge at Pb-site in both $\alpha$- and $\beta$-phases of CsPbI$_{3}$ and determined the relative stability of these lower Pb-content iodide perovskite compounds. With simultaneous substitutions at A and B-sites of the $\alpha$-phase, we intend to increase the stability of the desired phase of CsPbI$_{3}$ without altering the optoelectronic properties much. This is in cognizance of the finding that the contribution of entropy to free energy can be as high as 35 meV/f.u. with mixing at two sites in double perovskites \cite{wang2022entropy}. The electronic structure, effective mass, transition probability, absorption coefficient, and  theoretical efficiency (qualitative) are calculated and presented. Since mobility is an important figure of merit for semiconductors in developing efficient devices for PV applications, we determined the electron mobility at finite temperature using Feynman's polaron theory \cite{feynman1955slow, osaka1959polaron} and Rode's iterative method of solving the Boltzmann transport equation \cite{rode1970electron, rode1970electron1} to obtain an approximate estimate for these mixed configurations.

\section{Computational Details}
The vast alloy configuration space can be efficiently modeled using the cluster expansion (CE) method \cite{laks1992efficient, wolverton1994cluster}, as implemented in the Lattice Atomic Configuration Simulation (LACOS) package \cite{lacos}. This method uses a small set of first-principles data to train Hamiltonians with machine learning algorithms, which are then used to predict properties across the entire composition space. Solid solution alloy configurations of the $\alpha$- and $\beta$ phases of CsPbI$_{3}$ are obtained using DBmaker module of LACOS. For Ge mixing at the Pb site (in $\alpha$-phase), symmetrically distinct configurations are generated in 2$\times$2$\times$3 supercells (60 atoms) over the full composition range (x=0 to 1). Also, we consider configurations with Rb mixing at the Cs site for the $\alpha$-phase.  We found that supercells (2$\times$2$\times$2) with 40 atoms are sufficient to generate a manageable number of unique configurations for total energy density functional theory (DFT) calculations for simultaneous mixing at the A- and B-sites. For each structure, the volume is fully relaxed using an appropriate exchange-correlation functional. The total energy of alloy configurations are calculated by constructing the cluster expansion Hamiltonian, where the energy of an alloy configuration $\sigma$ \cite{das2023computational, ram2024exploring} is given by,
\begin{equation}
    E_{CE}(\sigma) = J_0 + \sum_\alpha J_\alpha \Bigg<\prod_{i\in\alpha} \sigma_i \Bigg> = J_0 + \sum_\alpha J_\alpha\Pi_\alpha
\end{equation}
Here, $\sigma$ = \{$\sigma_1$,$\sigma_2$,....,$\sigma_N$\} is an N-dimensional vector (N being the number of atomic sites in the crystal) that specifies which type of atom occupies the atomic site $i$. $J_\alpha$ is the effective cluster interactions (ECI) and $J_0$ represents the empty site energy. $\alpha$ stands for one, two or three-site clusters. $\Pi_\alpha$ are the cluster correlation functions.

From the total number of configurations, 80\% is used to train the CE model, while 20\% is reserved for testing. The least absolute shrinkage and selection operator (LASSO) was used to obtain the optimal $J_\alpha$ by minimizing the following function
\begin{equation}
   J_{opt} = \arg\min_J ||E_{DFT} - E_{CE}||_{2}^{2} + \lambda ||J||_{1}
\end{equation}
where $\lambda ||J||_{1}$ is the $L_1$ regularization term and $\lambda$ is a hyperparameter. The final model is selected by minimizing the root-mean-square-error (RMSE) and keeping the number of clusters as low as possible. The accuracy of the CE model is verified by comparing the CE-predicted energies with DFT-calculated energies of the hold-out set. The resultant RMSE is found to be 0.0017 and 0.004 eV/atom for one-site mixing (Ge at Pb-site) and two-site mixing (Rb at Cs-site and Ge at Pb-site) respectively (refer Figure S1 of supplemental material).

From the total energies, the enthalpy of mixing is calculated as
\begin{equation}
\Delta H_j = E_j - (1-x)E_{CsPbI_{3}} - xE_{CsGeI_{3}}
\end{equation}
and the configurational entropy per formula unit ($\Delta$S) and free energy of mixing ($\Delta$G) are obtained as
\begin{equation}
\Delta S = -k_B[xlnx + (1-x)ln(1-x)]
\end{equation}
\begin{equation}
 \Delta G = \Delta H_j - T\Delta S
\end{equation}
where T is the temperature in Kelvin. Here, we consider only the configurational entropy as it plays a dominant role (in general) compared to vibrational and electronic entropies. Similar approach is followed in previous reports on perovskites \cite{wang2022entropy, chen2021unified}.

DFT calculations are performed using Vienna ab-initio Simulation Package (VASP) \cite{kresse1993ab, kresse1994ab, kresse1996efficiency, kresse1996efficient} with wave functions described by the plane wave basis set and the ionic potential by the projector augmented wave (PAW) method \cite{blochl1994projector, kresse1999ultrasoft}. The energy cut-off of the plane wave basis set is 500 eV. The Perdew-Burke-Ernzerhof formalism of generalized gradient approximation (PBE) \cite{perdew1996generalized} is used to describe the exchange and correlation. The structures are optimized with the conjugate gradient algorithm until the residual force on each atom is less than 0.01 eV/{\AA} and energy is below 10$^{-6}$ eV. Since spin orbit coupling (SOC) has a sizeable effect on the electronic structure of heavy lead halides, it is used in conjunction with PBE. As presented in the Results and Discussion, we use PBE+U+SOC to determine the electronic structure of optimized solid solution models after benchmarking the band gap values of pure $\alpha$-CsPbI$_{3}$, rhombohedral CsGeI$_{3}$, and $\alpha$-RbPbI$_{3}$ with available experimental and theoretical data. 

We carried out density functional perturbation theory (DFPT) \cite{baroni2001phonons} calculations to obtain the dielectric response of electrons ($\epsilon_{\infty}$) and ions ($\epsilon_{ionic}$) at the PBE+U+SOC level. Although hybrid exchange-correlation functionals with SOC (HSE+SOC) give better band gaps with respect to PBE, we have not used it in the current work. This is because DFPT cannot be used in combination with hybrid functionals, which gives us the optical and static dielectric constants for transport calculations. The scattering rates and mobility are calculated using the Rode's iterative method of solving the Boltzmann transport equation as implemented in the AMMCR code \cite{mandia2019ab, chakrabarty2019semi, mandia2021ammcr}.

\section{Results and Discussion}
\subsection{Geometry, mixing energy and band gap}
Before moving on to partially substituted systems, we discuss the results of the unit cell used in our study. The calculated lattice constant of $\alpha$-CsPbI$_{3}$ is 6.39 {\AA} with PBE as functional, which is in agreement with other theoretical reports \cite{dong2021linear, su2021stability}. The difference with the experimental value is $\sim$ 1.6\%. Similarly, for $\alpha$-CsGeI$_{3}$ and $\alpha$-RbPbI$_{3}$, the lattice constants obtained are 6.00 and 6.38 {\AA}, respectively. The value of $\alpha$-RbPbI$_{3}$ is 0.07 {\AA} higher than that reported in an earlier study \cite{jong2018first}, but matches with Brgoch et al. \cite{brgoch2014ab} where van der Waals interaction is not considered. It should be noted that the experimental lattice constant value of $\alpha$-RbPbI$_{3}$ is not available. Also, the room temperature phase of CsGeI$_{3}$ is rhombohedral which is very close to cubic perovskite structure. The rhombohedral CsGeI$_{3}$ and $\alpha$-CsPbI$_{3}$ are used to obtain the appropriate U values for Ge-p and Pb-p orbitals by matching their band gap with the experimental values (details of which are given in supplemental material file). 

\subsubsection{B-site mixing: CsPb$_{1-x}$Ge$_{x}$I$_{3}$}
\begin{figure}
    \subfigure[]{\hspace{0.4cm}\includegraphics[trim=0mm 0mm 0mm 0mm,clip,scale=0.45]{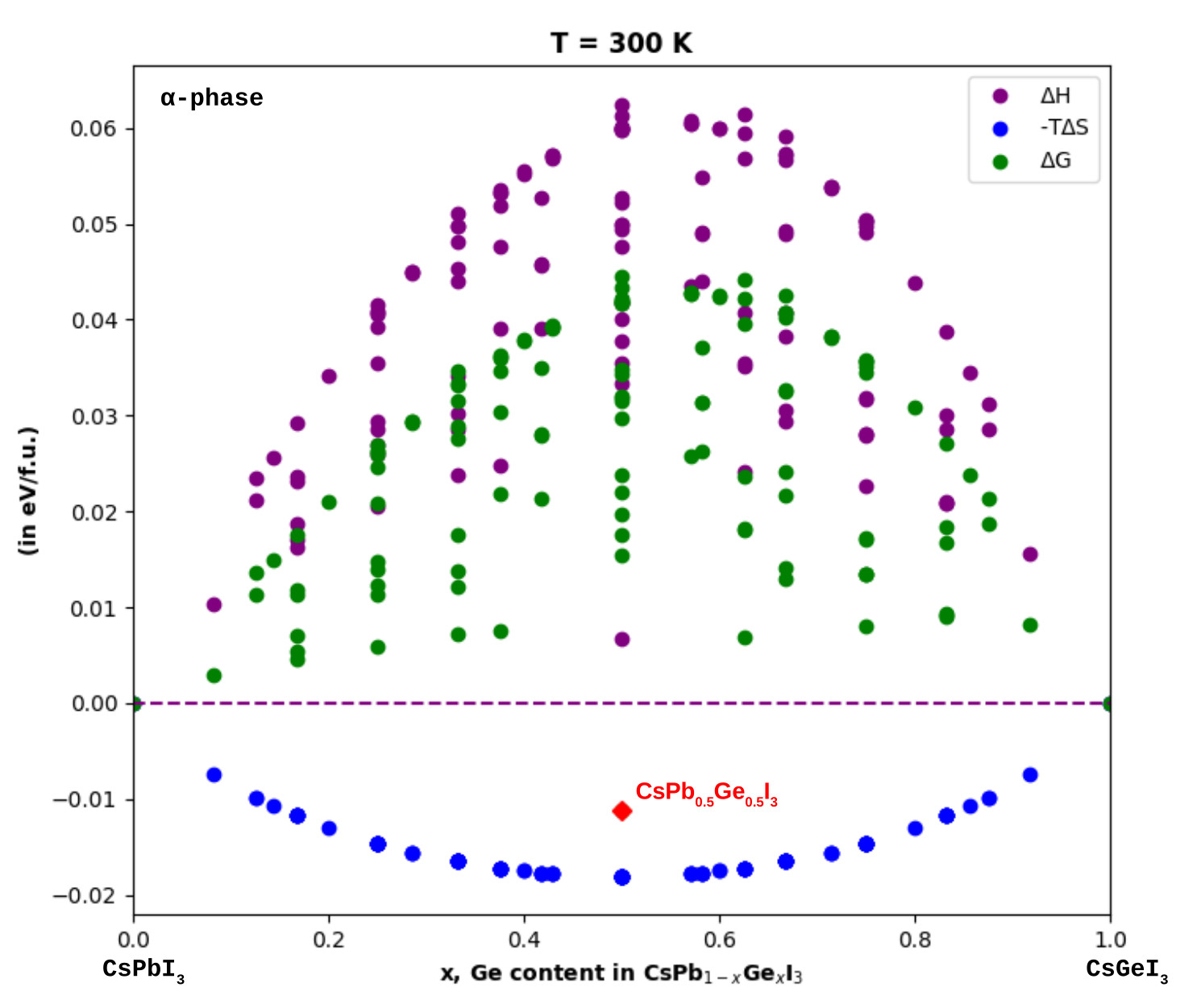}} \\
    \subfigure[]{\hspace{0.4cm}\includegraphics[trim=0mm 0mm 0mm 0mm,clip,scale=0.32]{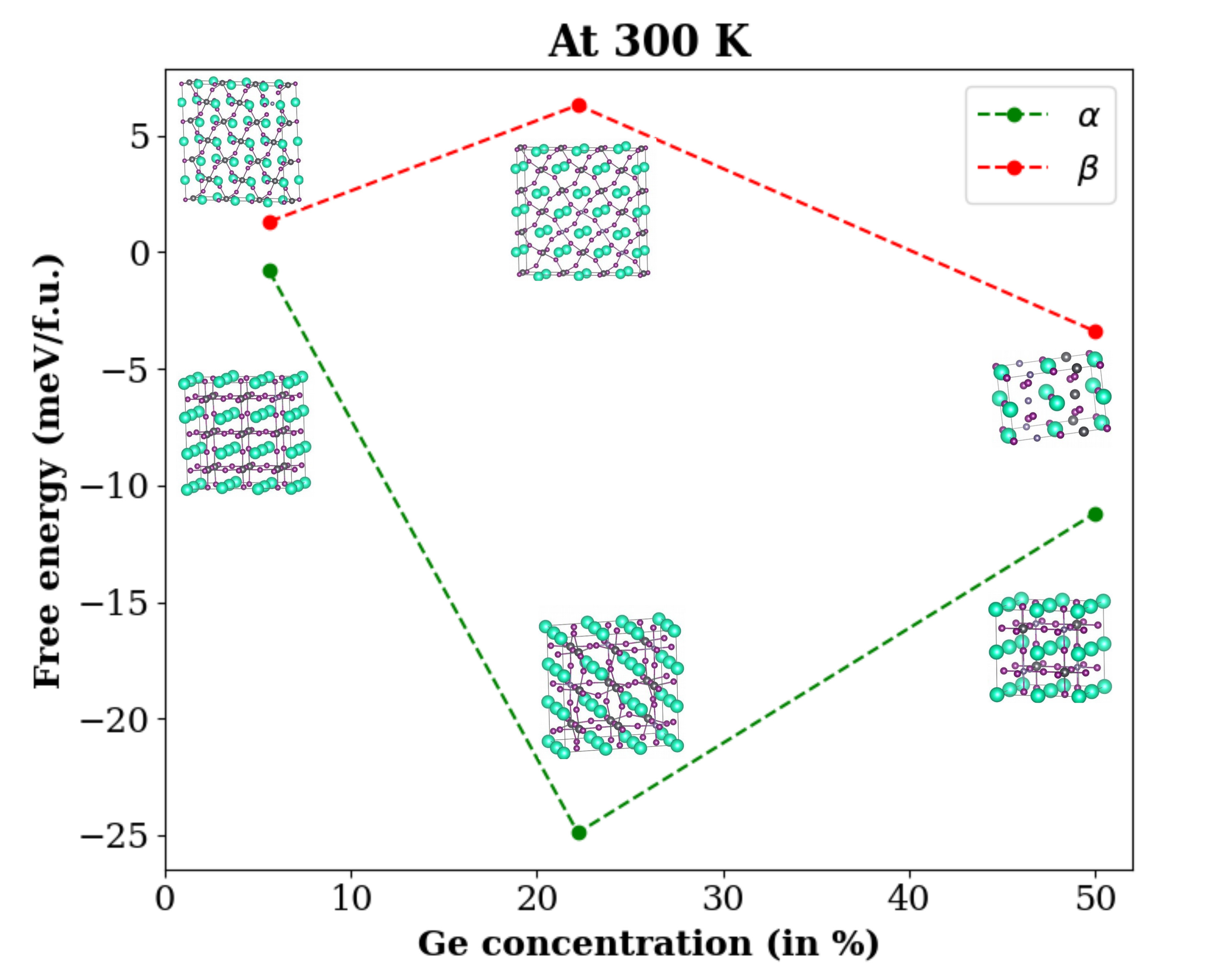}} \\
    \subfigure[]{\hspace{0.8cm}\includegraphics[trim=0mm 0mm 0mm 0mm,clip,scale=0.65]{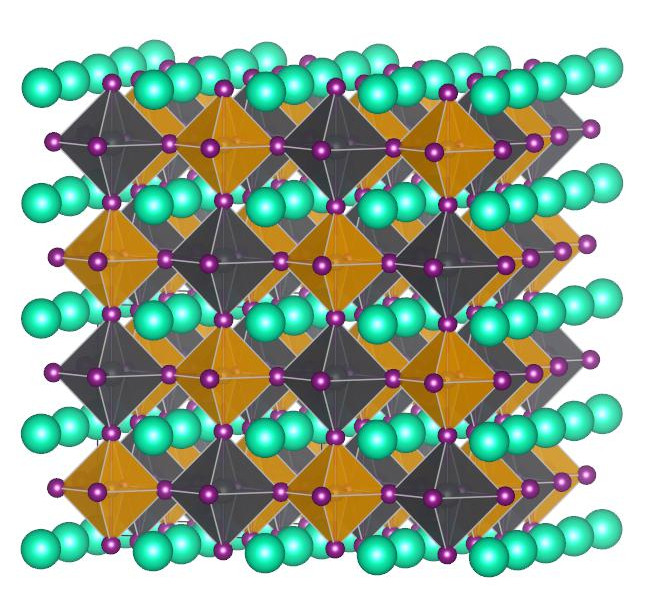}} 
    \subfigure[]{\hspace{0.1cm}\includegraphics[trim=0mm 0mm 0mm 0mm,clip,scale=0.70]{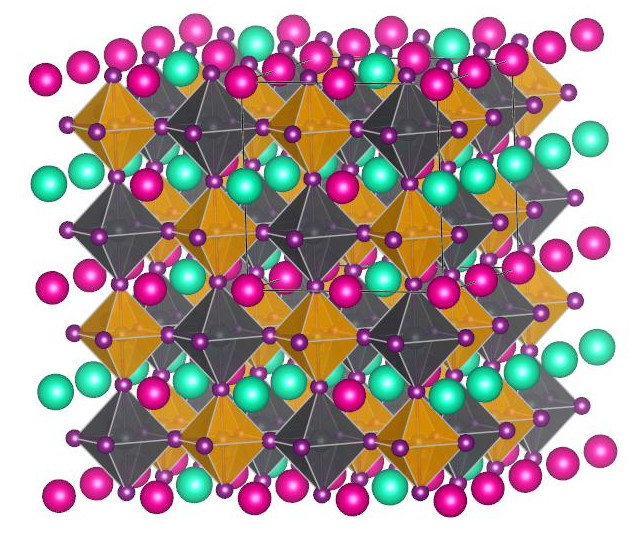}}
    \caption{(a) Mixing enthalpy and free energy at 300 K as a function of Ge content in $\alpha$-CsPb$_{1-x}$Ge$_{x}$I$_{3}$ (the red diamond corresponds to the Gibb's free energy of CsPb$_{0.5}$Pb$_{0.5}$I$_{3}$). (b) Gibb's free energy of $\alpha$-CsPb$_{1-x}$Ge$_{x}$I$_{3}$ and $\beta$-CsPb$_{1-x}$Ge$_{x}$I$_{3}$ at 300 K as a function of Ge content. The arrangement of atoms in configurations (c) $i$ and (d) $iv$. Yellow octahedra are centered at Ge and black octahedra are centered at Pb. Green, magenta, and purple balls correspond to Cs, Rb, and I respectively.}
    \label{G_struct_plot}
\end{figure}

In the $\alpha$ and $\beta$ phases of CsPbI$_{3}$, Ge alloying is carried out using cluster expansion approach. The mixing enthalpy and free energy of $\alpha$-CsPbGeI$_{3}$ at 300 K as a function of Ge content is presented in Fig \ref{G_struct_plot}(a). The respective plot for $\beta$-CsPbGeI$_{3}$ is given in Figure S2(a). From the determination of effective cluster interactions (ECI), minimum energy configurations are predicted by the Monte-Carlo module of LACOS for the three alloying concentrations of interest. They are 5.6 (minor), 22.2 and 50\% (high-concentration) alloying. DFT calculations are further carried out to obtain the total energy. The Gibb's free energy as a function of Ge for both $\alpha$ and $\beta$-phases are given in Figure \ref{G_struct_plot}(b). In the $\alpha$-phase, at all three concentrations, the free energy takes negative value. But for $\beta$-phase, free energy is negative only at 50\% alloying. Therefore, it can be concluded that it is $\alpha$-phase which is energetically more stable with respect to $\beta$-phase for the three compositions with lower lead content. These structures have tetragonal (123-P4/mmm), monoclinic (6-Pm) and cubic (225-Fm$\bar{3}$m) symmetry respectively. Though at 22.2\% alloying, the configuration has large negative Gibb's free energy, but the bond angles deviate a lot from 180$^{\circ}$ (may be a non-perovskite). The angular deviation for some exceeds 5\%. This is not so for 5.6 and 50\% compositions. A quick inspection of their band gap at PBE level show that the values are 1.43, 1.41 and 1.06 eV for 5.6, 22.2 and 50\% alloying respectively. The optimal range of band gap to achieve a Shockley-Queisser efficiency $\sim$25\% is 0.9-1.6 eV \cite{ju2017lead} and for the three compositions, this criterion is satisfied.

Among these, the representative configuration of CsPb$_{0.5}$Ge$_{0.5}$I$_{3}$ (called $i$ from this point) has highest symmetry. This structure is formed when Ge and Pb atoms occupy alternate positions in the sublattice (i.e. alternate arrangement of PbI$_{6}$ and GeI$_{6}$ octahedra in a rock-salt face centered cubic structure) in all three directions given in \ref{G_struct_plot}(c). It is crystallographically identical to ordered double perovskites (elpasolite structure) with spacegroup Fm$\bar{3}$m and chemical formula Cs$_{2}$PbGeI$_{6}$. As can be seen, this structure is stabilized by entropy and could correspond to the stable ordered phase of CsPb$_{0.5}$Ge$_{0.5}$I$_{3}$. It has a Gibb's free energy of -11.2 meV/f.u. (at 300 K). It should be noted that the vibrational entropy is not taken into account as it is considered to have negligible effect. The ionic radii of Pb and Ge are 1.19 and 0.73 \AA{} respectively and the mismatch is around 0.46 \AA{}. There is quite a difference in Gibb's free energy (or energy of mixing) of this ordered structure and other configurations with the 50-50 composition, which could be attributed to the alternate arrangement of Pb and Ge minimizing the internal strain arising from the ionic radii mismatch between them. This difference in energy is as large as 23 meV/f.u. obtained by searching all supercells with a maximum of 40 number (n) of atoms (we are not looking at supercells with n $>$ 40, as transport calculations become cumbersome). Also, the relative strength of Pb-I (3.20 \AA{}) and Ge-I (3.00 \AA{}) bonds in configuration $i$ differ, that is, the bonding strength of Pb-I bonds is found to be stronger than that of Ge-I bonds. A previous study based on the cluster expansion method reported a similar configuration with negative formation energy, which concluded that mixed divalent cations at B site stabilize the perovskite structure when the A site is occupied by Cs$^{+}$ ion \cite{yamamoto2017structural}. 

Interestingly, this alternate arrangement of Pb- and Ge-centered octahedra could be a condition for high resistance to oxidation of Ge. As reported in the case of ASnI$_{3}$ and APb$_{0.5}$Sn$_{0.5}$I$_{3}$,  perovskites, the formation of SnI$_{4}$ along with SnO$_{2}$ requires the cooperative action of several SnI$_{6}$ octahedra lying adjacent to each other \cite{leijtens2017mechanism}. When a Sn$^{2+}$, that is being oxidized has Pb$^{2+}$ in the adjacent octahedra, the cooperative mechanism that pave the way for oxidation to 4+ state becomes less favorable. Instead, it is forced to take the pathway that requires more Sn-I and Pb-I bonds to break to form I$_{2}$, which involves high activation energy and thereby preventing it. Hence, by incorporating smaller-sized Ge cations in alternate positions with Pb, it may be possible to slow the oxidation reaction just as in the case of Sn.

We look at other configurations with lower symmetry (i.e. higher energy) at 50\% alloying. There are two configurations with the same mixing enthalpies (hence the same free energies), which are found to have tetragonal symmetry. They have a $\Delta$G of 12 meV/f.u. (at 300 K). The difference in total energy with respect to configuration $i$ is $\sim$28 meV/f.u. The first configuration with tetragonal symmetry is a 2$\times$2$\times$2 supercell with 40 atoms and the other configuration is a 1$\times$2$\times$2 supercell with 20 atoms. For further calculations, we consider the second configuration with a smaller number of atoms (to reduce computational cost) named $ii$, here onward. In the yz plane, this structure has alternate ordering of the PbI$_{6}$ and GeI$_{6}$ octahedra, but in x-direction, alternate rows of Pb- and Ge-centered octahedra are present (given in the supplemental material, Figure S2).

The corresponding lowest energy structure of $\beta$-CsPb$_{0.5}$Ge$_{0.5}$I$_{3}$ (called $iii$) has all Ge-centered octahedra lying adjacent to each other, whereas Pb-centered octahedra aggregate together unlike $\alpha$ phase. The atomic arrangement of this configuration is shown in Figure S2(b). Here, the probability of oxidation of Ge to the 4+ state is high. 

\subsubsection{A- and B-site mixing: Cs$_{1-y}$Rb$_{y}$Pb$_{1-x}$Ge$_{x}$I$_{3}$}
Since configuration $i$ is stabilized by entropy, it may not be a bad idea to use the large configurational entropy of multi-element alloys to offset the positive mixing-enthalpy barrier. Therefore, we consider two-site mixing in $\alpha$-CsPbI$_{3}$ to improve its thermodynamic stability at finite temperature. In a previous report \cite{jong2018first}, Cs and Rb was found to have good miscibility in their solid solution (Cs$_{1-x}$Rb$_{x}$PbI$_{3}$). It was shown that for the entire range of Rb, solid solutions were more stable than their constituents at 300 K. Although we have carried out total energy calculations for the full range of x and y in Cs$_{1-y}$Rb$_{y}$Pb$_{1-x}$Ge$_{x}$I$_{3}$, here we discuss the composition for which the contribution of configurational entropy is highest (i.e. x=0.50 and y=0.50). For Cs$_{0.5}$Rb$_{0.5}$Pb$_{0.5}$Ge$_{0.5}$I$_{3}$, the lowest energy configuration has tetragonal symmetry with alternate ordering of PbI$_{6}$ and GeI$_{6}$ octahedra in x-, y-, and z-directions (called as $iv$). The ordering observed in configuration $i$ is retained in $iv$ (presented in Figure \ref{G_struct_plot}(d)). The calculated mixing enthalpy $\Delta$H$_{max}$ and change in Gibb's free energy $\Delta$G$_{max}$ as a function of entropy contribution T$\Delta$S (at 300 K) is given in supplemental material. We have followed the method used in reference\cite{wang2022entropy}, to calculate mixing enthalpy using four reaction pathways. With 50\% alloying of Rb at the Cs-site, the free energy changes to -20.5 meV/f.u. from -11.2 meV/f.u. at 300 K (entropy stabilization).

\subsection{Mechanical stability}
We estimated the elastic constants of configurations $i$ and $iv$ to determine their mechanical stability. Since configuration $i$ has cubic symmetry, there are three independent elastic constants: C$_{11}$, C$_{12}$, and C$_{44}$. They satisfy Born stability criteria for cubic crystals (C$_{11}$ $>$ 0, C$_{44}$ $>$ 0, C$_{11}$ - C$_{12}$ $>$ 0 and C$_{11}$ + 2C$_{12}$ $>$ 0), which implies that they are mechanically stable at zero pressure. The elastic constants are C$_{11}$ = 36.47, C$_{12}$ = 6.01, C$_{44}$ = 5.59 GPa for configuration $i$. The corresponding elastic constants for $\alpha$-CsPbI$_{3}$ are C$_{11}$ = 34.84, C$_{12}$ = 4.73, C$_{44}$ = 3.66 GPa. There is a slight increase in the values of elastic constants compared to $\alpha$-CsPbI$_{3}$. Similarly, there is an increase in the bulk modulus (16.16 GPa) with respect to $\alpha$-CsPbI$_{3}$ (14.76 GPa) (as in the Voigt-Reuss-Hill approximation). The increase in bulk modulus with Ge alloying could mean a decrease in lattice softness. The configuration $iv$ has tetragonal symmetry and has six independent elastic constants. They are C$_{11}$ = 37.02, C$_{12}$ = 5.95, C$_{13}$ = 5.99, C$_{33}$ = 36.97, C$_{44}$ = 4.26 and C$_{66}$ = 4.77 GPa. This configuration too, meets the elastic stability criteria (C$_{11}$ $>$ $\vert$C$_{12}\vert$, 2C$_{13}^{2}$ $<$ C$_{33}$(C$_{11}$ + C$_{12}$) and C$_{44}$ $>$ 0).

\subsection{Electronic structure and Optical Properties}
\begin{figure}
	\begin{center}
    \subfigure[]{\hspace{-0.2cm}\includegraphics[scale=0.60]{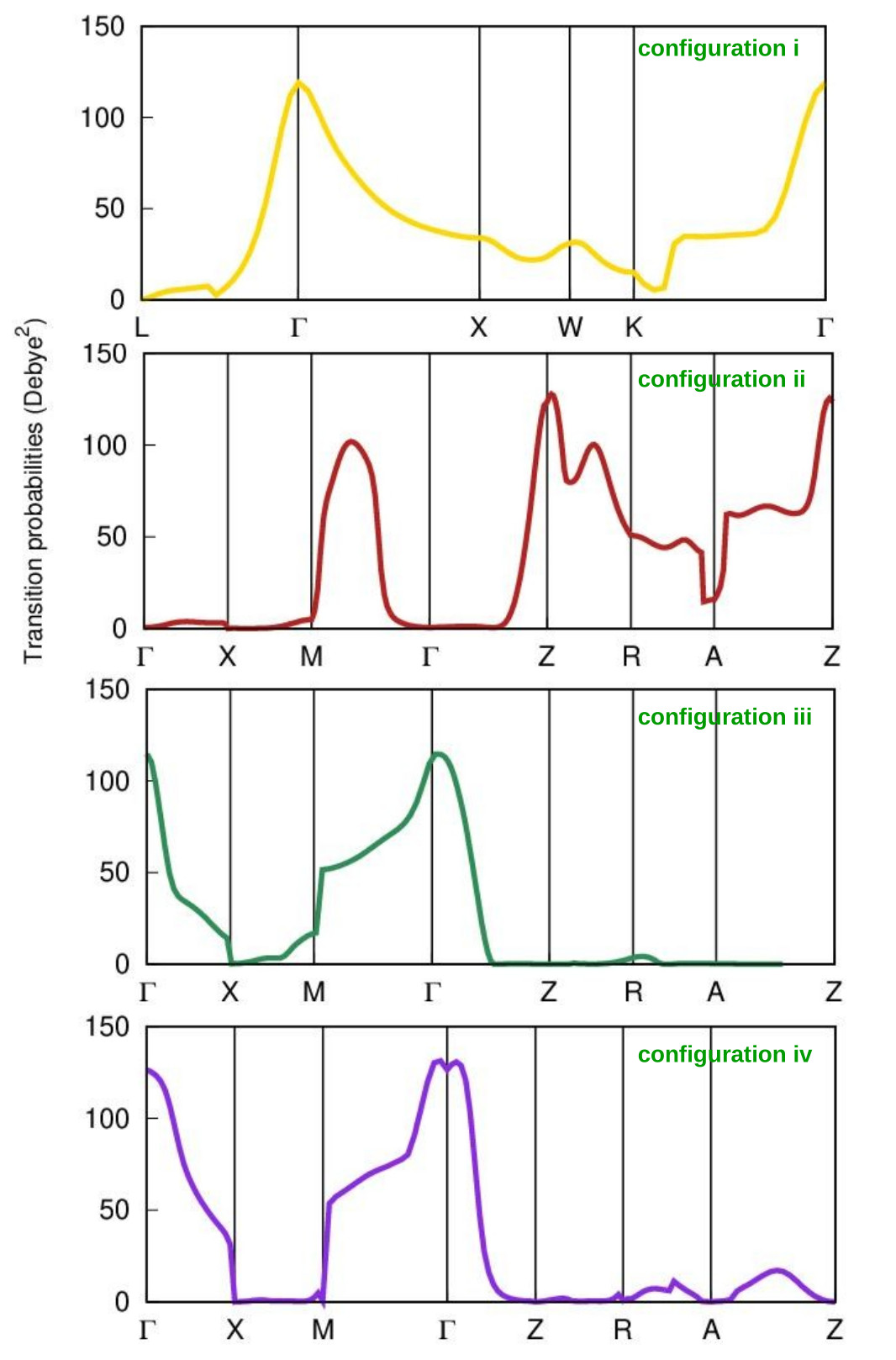}} 
    \subfigure[]{\hspace{-0.2cm}\includegraphics[scale=0.35]{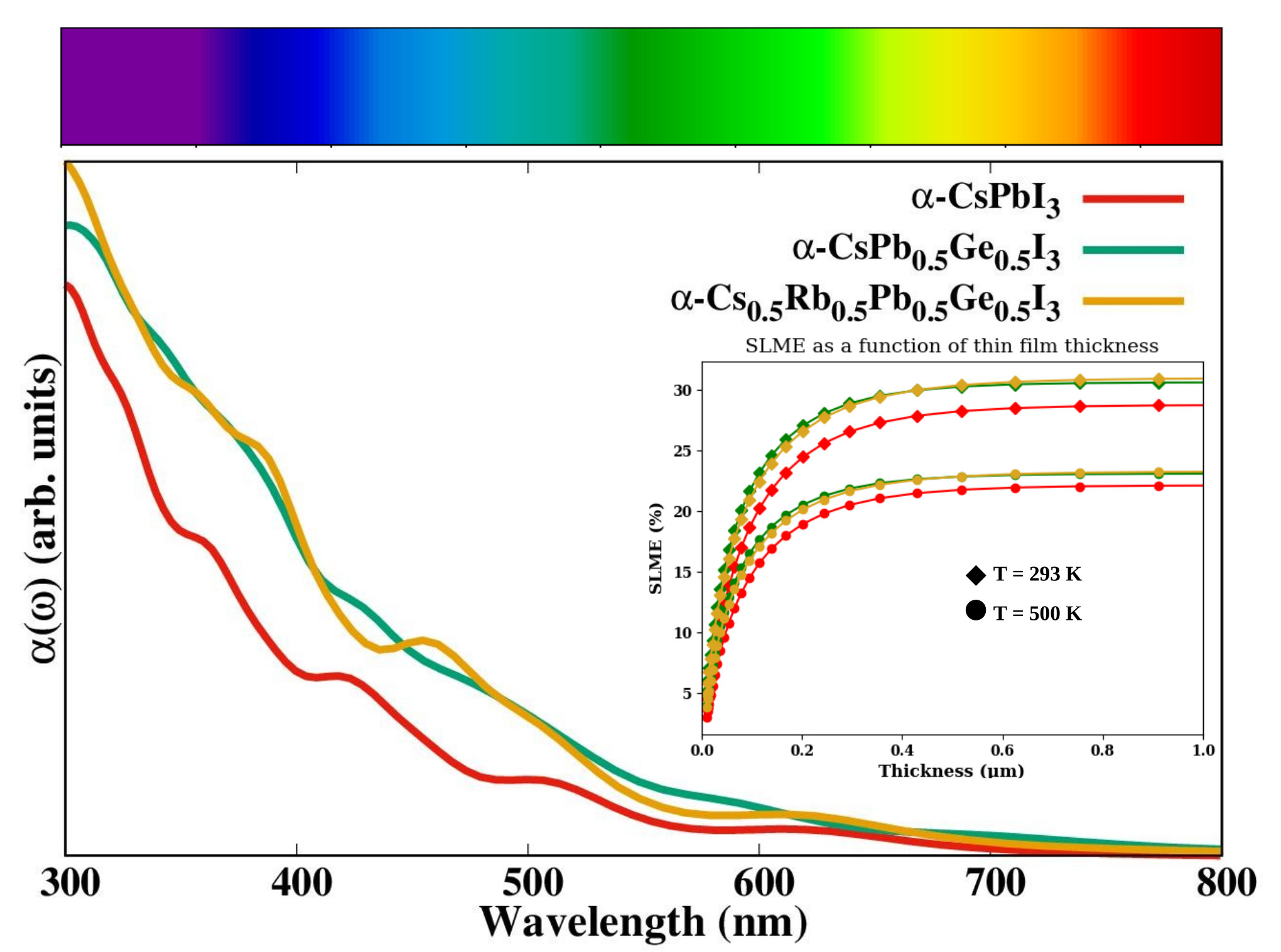}} 
    \caption{(a) Transition matrix amplitudes of configurations $i$, $ii$, $iii$ and $iv$ obtained from PBE+U+SOC band structure calculation. (b)Absorption spectra of $\alpha$-CsPbI$_{3}$, $\alpha$-CsPb$_{0.5}$Ge$_{0.5}$I$_{3}$ (configuration $i$), and $\alpha$-Cs$_{0.5}$Rb$_{0.5}$Pb$_{0.5}$Ge$_{0.5}$I$_{3}$ (configuration $iv$). The inset shows the spectroscopic limited maximum efficency (SLME) for the same. The results of $\alpha$-CsPbI$_{3}$ are shown for comparison.}
    \label{abs_SLME}
    \end{center}
\end{figure}

The electronic band structure of the ordered configuration $i$ has been determined considering the primitive cell with 10 atoms. In a typical PBE calculation, the band gap obtained is 1.06 eV and the conduction band minimum (CBM) is found to be triply degenerate. As is common among perovskites, the valence band maximum (VBM) of this configuration exhibits only a single value. SOC correction is included because of the presence of elements such as Pb and Ge, with the effect being greater for heavier element Pb compared to that of Ge. With SOC, the degenerate unoccupied Pb p-orbitals split, causing a reduction in the band gap to 0.34 eV. We used DFT+U to overcome the underestimation in the band gap caused by SOC. The appropriate values of U for p-orbitals of Pb and Ge [U$_{p}$(Pb) = 12.5 and U$_{}$(Ge) = 9.5 eV] are calculated by matching the band gap of $\alpha$-CsPbI$_{3}$ and rhombohedral CsGeI$_{3}$ with experimental values (further details are provided in the supplemental material). It is a direct band gap semiconductor with E$_{g}$ = 1.58 eV (value obtained in PBE+U+SOC calculation). The formation of a direct gap is an advantage compared to other lead-free double perovskites \cite{meng2017parity}. The CBM is mainly derived from the Pb-p states and lies at the $\Gamma$ point. The VBM consists of Pb-s and I-p orbitals. The electron orbitals of Cs lie away from the Fermi level, deep in the valence band contributing little to the electron transitions due to photovoltaic effect. Replacement of Pb with the isovalent Ge or Sn results in an electronic structure with features similar to the CsPbI$_{3}$ band structure. In a previous study by Meng et al. \cite{meng2017parity}, they showed that CsGeI$_{3}$ and CsPbI$_{3}$ exhibits larger band gap than CsSnI$_{3}$, which is attributed to the 4s and 6s energy levels of Ge and Pb lying closer to each other than 5s energy level of Sn. The configuration $i$ does not exhibit a parity-forbidden transition at $\Gamma$ point as observed in the large transition probability (given in Figure \ref{abs_SLME}(a)), calculated from the dipole transition matrix elements (explained in the following paragraphs). 

\begin{figure}
    \begin{center}
    \subfigure[]{\hspace{0.0cm}\includegraphics[trim=0mm 16mm 0mm 0mm,clip,scale=0.25]{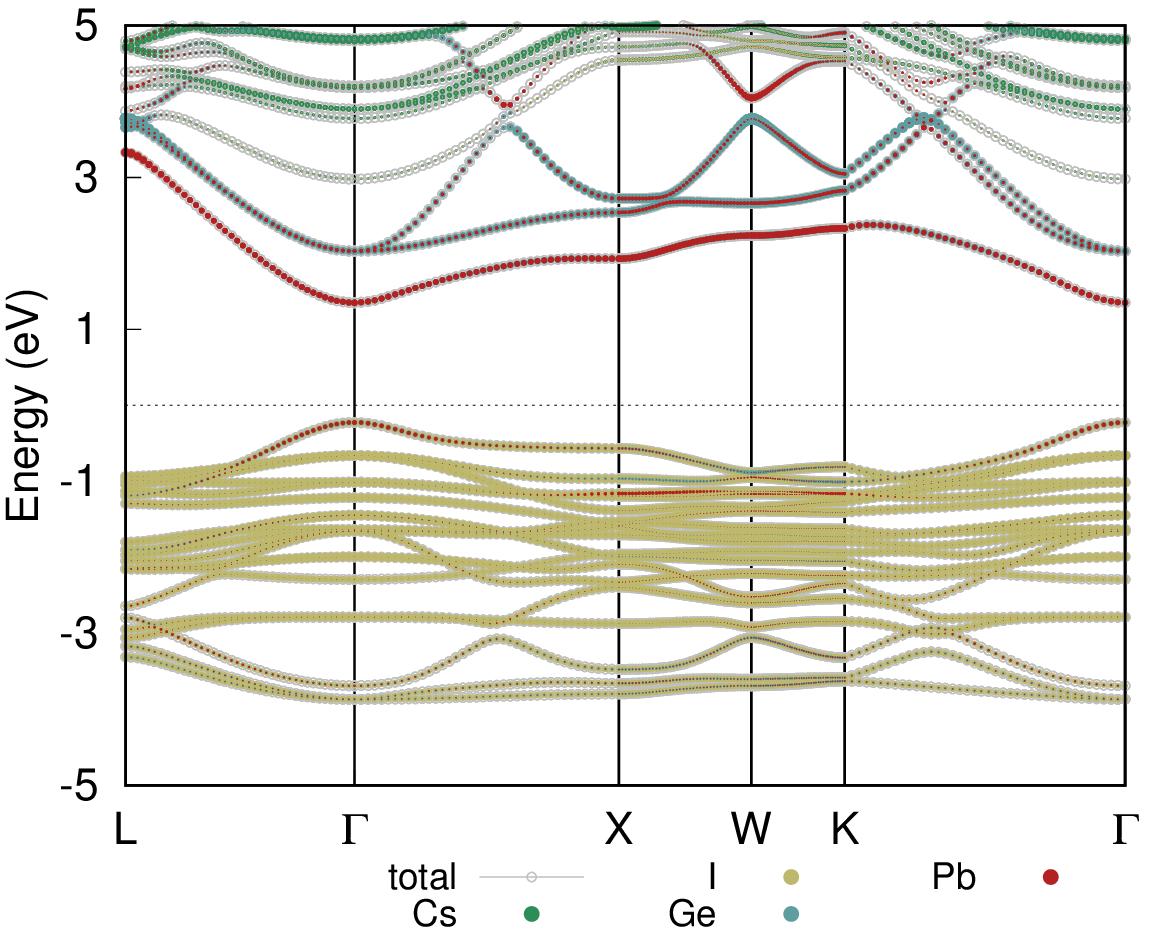}} \\
    \subfigure[]{\hspace{0.0cm}\includegraphics[trim=0mm 16mm 0mm 0mm,clip,scale=0.25]{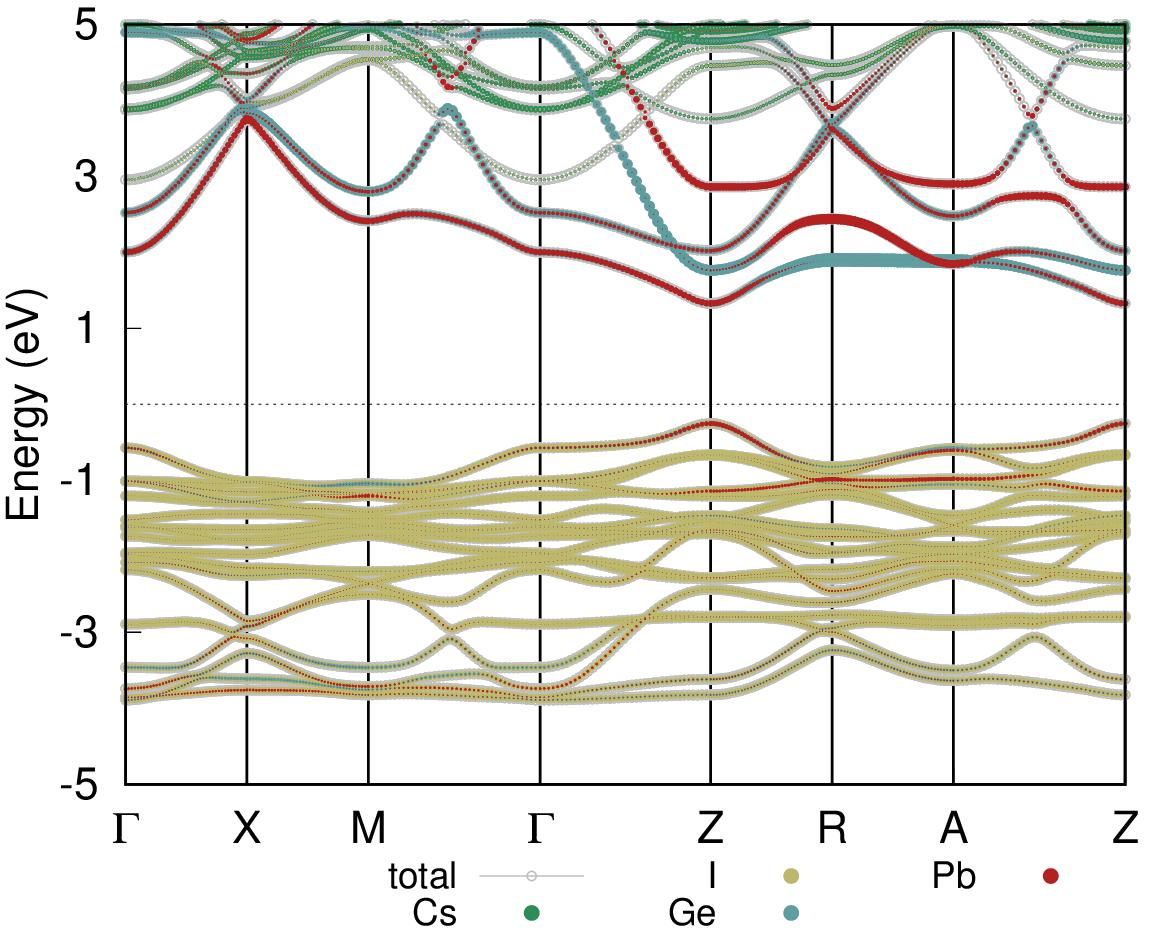}} \\
    \subfigure[]{\hspace{0.0cm}\includegraphics[trim=0mm 16mm 0mm 0mm,clip,scale=0.25]{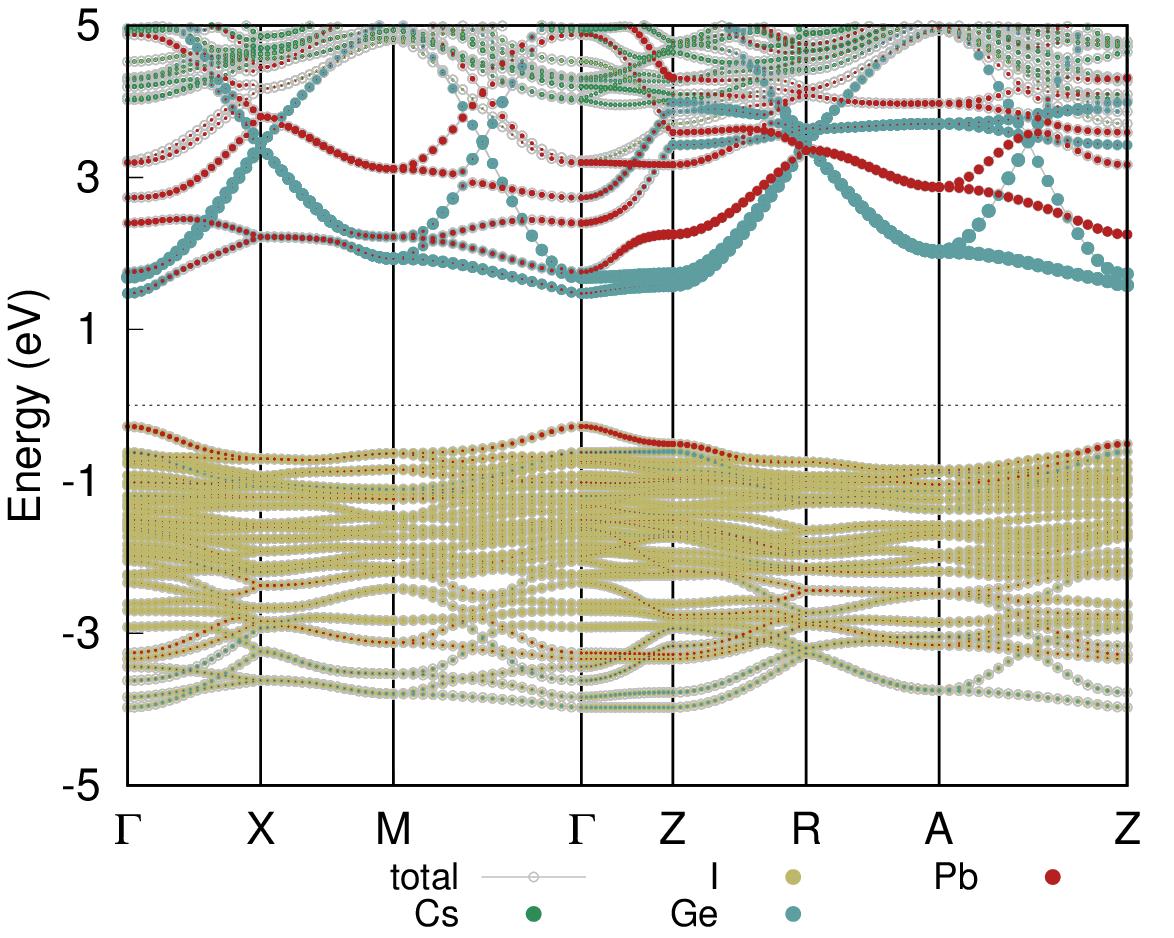}} \\
    \subfigure[]{\hspace{0.0cm}\includegraphics[trim=0mm 0mm 0mm 0mm,clip,scale=0.25]{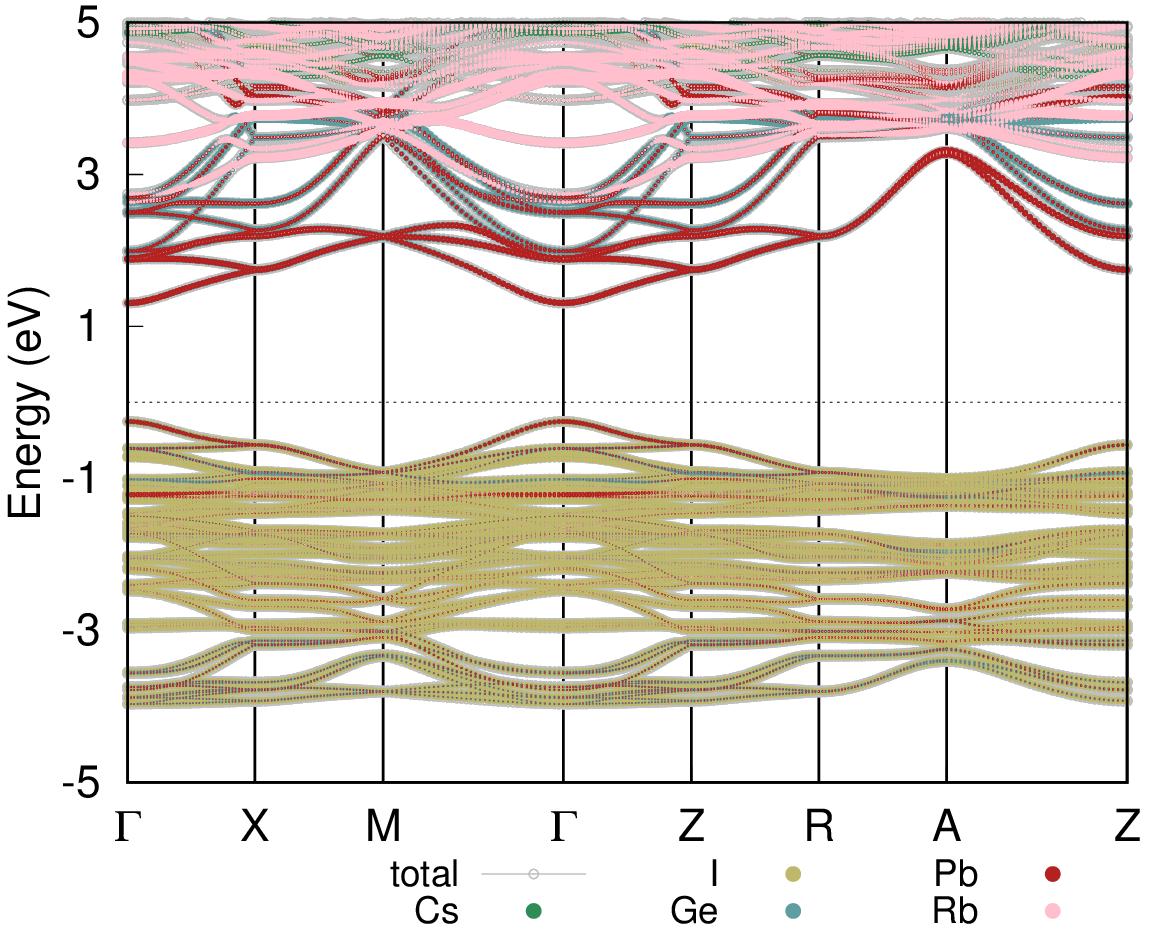}}
    \caption{Element resolved band structure of configurations a) $i$ (cubic $\alpha$-CsPb$_{0.5}$Ge$_{0.5}$I$_{3}$), b) $ii$ (tetragonal $\alpha$-CsPb$_{0.5}$Ge$_{0.5}$I$_{3}$), c) $iii$ ($\beta$-CsPb$_{0.5}$Ge$_{0.5}$I$_{3}$) and d) $iv$ (Cs$_{0.5}$Rb$_{0.5}$Pb$_{0.5}$Ge$_{0.5}$I$_{3}$) along the corresponding high symmetry paths. For configurations $i$ and $ii$, primitive cells are taken into consideration.}
    \label{bnd}
    \end{center}
\end{figure}
The projected band structure of the primitive cell of configuration $ii$, with the same chemical composition as $i$ (i.e. CsPb$_{0.5}$Ge$_{0.5}$I$_{3}$) is given in Figure \ref{bnd}(b). It has a direct band gap of 1.58 eV. Although there is a decrease in symmetry, the band gap value is maintained. For the $\beta$ phase of CsPb$_{0.5}$Ge$_{0.5}$I$_{3}$ (i.e. $iii$), the band gap obtained using similar method is 1.76 eV with VBM and CBM lying at $\Gamma$. Therefore, the $\alpha$-phase has a better band gap for PV application. The value for the $\beta$-phase is slightly above the optimal range. The configuration $iv$ (i.e. Cs$_{0.5}$Rb$_{0.5}$Pb$_{0.5}$Ge$_{0.5}$I$_{3}$) with Rb- and Ge-mixing has a band gap of 1.56 eV from PBE+U+SOC calculations. The gap is direct in nature with the band extremas lying at $\Gamma$ point.

In some cases, the inversion symmetry of metal halide and double perovskites leads to parity-forbidden transitions, which can affect optoelectronic properties \cite{meng2017parity}. To examine this, we calculated the transition matrix amplitudes between the valence band and the conduction band for the four configurations ($i$, $ii$, $iii$, and $iv$) and they are presented in Figure \ref{abs_SLME}. The transition probability is highest at the $\Gamma$ point for the configuration $i$, which implies that it has a good optical absorption coefficient for photons with energies close to the band gap, an ideal condition for thin-film PV applications. At k-points away from the $\Gamma$, the symmetries of the valence and conduction bands change and the transition probability decreases (shown in Figure \ref{abs_SLME}(a)). It has a forbidden transition at the L point as the amplitude is zero. Even for the other three configurations, the transition probability is highest between VBM and CBM, which decreases at k-points away from it. In configurations $iii$ and $iv$, transitions along the Z-R direction are forbidden, which is not the case for configuration $ii$.

We have investigated the optical properties of the configurations $i$ and $iv$ by calculating the absorption coefficient as a function of wavelength. For comparison purposes, the absorption coefficient of $\alpha$-CsPbI$_{3}$ is also determined. Both configurations $i$ and $iv$, show an increase in optical activity in the visible region compared to $\alpha$-CsPbI$_{3}$, which can be attributed to the reduction in the band gap caused by Ge substitution at Pb-site and Rb-substitution at Cs-site. In other words, band gap narrowing induces a red shift in the absorption spectrum. These structures have good optical properties as a result of direct band gaps and no parity-forbidden transitions. The spectroscopic limited maximum efficiency (SLME) is also determined by using band gap and absorption coefficient calculated from first principles, and thickness as inputs. The Shockley-Queisser limit provides a relation between the band gap and the maximum theoretical efficiency of an absorber material in a single junction solar cell \cite{shockley2018detailed}. Yu and Zunger introduced SLME by expanding on the work of Shockley and Queisser \cite{yu2012identification}. This is implemented in the SL3ME package by L. Williams \cite{SL3ME}, which is used to obtain the qualitative trend of efficiency of mixed perovskite compositions when used as an absorber. Here, the standard AM1.5G solar spectrum is used \cite{amsolar}. Since all three configurations considered have direct band gaps, it is assumed that radiative recombination is the only recombination process for these materials. For indirect band gap materials, nonradiative recombination could be significant. The calculated SLME curves are almost constant for thicknesses larger than 0.8 $\mu$m. Both configurations $i$ and $iv$ have similar light harvesting performances at 293 and 500 K (shown in the inset of Figure \ref{abs_SLME}(b)) and are superior to $\alpha$-CsPbI$_{3}$. This shows that these mixed configurations are good candidate materials for highly efficient solar energy conversion. The SLME values of lead-free direct gap double perovskites, Cs$_{2}$InSbCl$_{6}$ and Cs$_{2}$InBiCl$_{6}$ reach 31 and 30\% respectively, at 1 $\mu$m thickness, surpassing 29\% of CH$_{3}$NH$_{3}$PbI$_{3}$ \cite{zhao2017design}. At 50 $\mu$m thickness, the SLME of Si (E$_{g}$ = 1.12 eV), GaAs (E$_{g}$ = 1.42 eV), and CsGeI$_{3}$  (E$_{g}$ = 1.53 eV) is evaluated to be 26.7, 29.1, and 28.37\% respectively \cite{chelil2023insights}.

The effective mass has been calculated for the configurations of interest using open source code $effmass$ \cite{whalley2018effmass}. For configuration $i$, the electron (hole) effective mass is 0.07 (-0.07) and 0.17 (-0.21) m$_{0}$ with PBE+SOC and PBE+U+SOC calculations. At the PBE level, the value for heavy electrons is as high as 0.76 m$_{0}$. This is consistent with the high value of 0.84 m$_{0}$ for $\alpha$-CsPbI$_{3}$ -  which is expected to limit electron transport and can lead to incorrect results \cite{su2021stability}. Hence, we consider the effective mass obtained using PBE+U+SOC in our further calculations. For the other two configurations $ii$ and $iv$, the corresponding electron effective masses are 0.18 and 0.17 m$_{0}$ respectively. The effective mass of hole is also in the same range as $i$, with it slightly higher for $ii$ (0.23 m$_{0}$), which implies a lower mobility.

\subsection{Scattering rate and mobility}
For the development of highly efficient devices, it is pertinent to know how charge carriers interact with phonons. Due to the polar nature of halide perovskites, charge carriers induce lattice distortion via Coulomb interactions, which are described by quasi-particles called polarons. We have calculated the dimensionless Fr{\"o}hlich electron-phonon coupling constant $\alpha_{el-ph}$ \cite {frohlich1954electrons, frost2017calculating} for $i$, $ii$ and $iv$ configurations using \cite{polaronf}. The expression for $\alpha_{el-ph}$ is given by
\begin{equation}
    \alpha_{el-ph} = \frac{1}{4\pi\epsilon_{0}}\frac{1}{2}\left(\frac{1}{\epsilon_{\infty}}-\frac{1}{\epsilon_{s}} \right)\frac{e^{2}}{\hbar\omega_{eff}} \left(\frac{2m_{b}^{*}\omega_{eff}}{\hbar} \right)^\frac{1}{2} 
\end{equation}
where $\epsilon_{\infty}$ and $\epsilon_{s}$ are the optical and static dielectric constants; m$_{b}^{*}$ is the electron band effective mass; $\omega_{eff}$ is the effective LO-phonon frequency; $e$, $\hbar$ and $\epsilon_{0}$ are the charge, reduced Planck constant and permittivity of free space respectively. The parameters used in our calculations are given in Table \ref{feymanP_parameters}. $\alpha_{el-ph}$ gives the strength of the electron-phonon coupling. Inorganic halide perovskites have $\alpha_{el-ph}$ coupling in the range 1 to 2 whereas hybrid halide perovskites have in the range 2 to 3 due to the additional dielectric response of the molecular cation \cite{frost2017calculating}. The effective phonon frequency is used to map the effect of scattering from the full phonon band structure to a single phonon frequency. This frequency calculated in the harmonic approximation can shift at finite temperatures due to inclusion of anharmonic effects. In the above expression, instead of the effective LO-phonon frequency, we have used the frequency of the most polar optical mode at $\Gamma$ (an approximation) obtained by calculating the mode effective charge (Born effective charges and phonon eigen vectors from the DFPT calculation). For the three configurations of interest, phonon spectra have imaginary modes at 0 K, which is common for perovskites. Some of the optical modes have negative frequencies and it is beyond the scope of the present work to calculate the renormalized phonon frequencies with the inclusion of higher-order interatomic force constants. The respective values of $\alpha_{el-ph}$ for the configurations $i$, $ii$, and $iv$ are 1.99, 2.09 and 1.86, which emphasize the formation of large polarons ($\alpha_{el-ph}$ $<$ 6) in these compositions. Large polarons in lead halide perovskites are expected to protect charge carriers against scattering and recombination \cite{zhu2016screening, zhu2015charge, qian2023photocarrier, lafuente2024topological}. Based on the Fr{\"o}hlich electron-phonon coupling model, large static dielectric constant of the halide perovskite causes long-range interactions (large polarons to form) and the net repulsion between oppositely charged large polarons will impede recombination \cite{emin2018barrier}. In addition to low recombination rate, other exceptional properties like high power conversion efficiency have been attributed to this large polaron formation. Most studies report the formation of large polarons in hybrid perovskites but Miyata et al. showed that they are formed by the deformation of the BX$_{3}^{-}$ frameworks in perovskite irrespective of the cation type (Cs or CH$_{3}$NH$_{3}$) \cite{miyata2017large}.  The large polaron takes longer to form in all-inorganic CsPbBr$_{3}$ (0.7 ps) compared to its hybrid counterpart (CH$_{3}$NH$_{3}$PbBr$_{3}$, 0.3 ps). The commonly found double halide perovskites like Cs$_{2}$AgBiBr$_{6}$, Cs$_{2}$AgBiCl$_{6}$, Cs$_{2}$AgInCl$_{6}$ have $\alpha_{el-ph}$ values of 2.54, 2.82 and 1.99 respectively \cite{steele2018giant, manna2020lattice}. For the $i$, $ii$ and $iv$ configurations, $\alpha_{el-ph}$ is in a similar range. The small values of the coupling constant could be a direct consequence of small electron effective mass (given in Table \ref{feymanP_parameters}) in the conduction band.

\begin{table}[h]
\small
 \caption{Parameters used in the Feynman polaron model. Frequency ($\omega$) is in THz and electron effective mass (m$_{b}^{*}$) is in units of bare electron mass.}
 \label{feymanP_parameters}
 \begin{tabular*}{0.48\textwidth}{@{\extracolsep{\fill}}lllll}
   \hline
   Material & $\epsilon_{\infty}$ & $\epsilon_{s}$ & $\omega$ & m$_{b}^{*}$\\
   \hline
   Configuration-$i$ & 4.83 & 30.48 & 4.29 & 0.17  \\
   Configuration-$ii$ & 4.85 & 37.29 & 4.35 & 0.18 \\
   Configuration-$iv$ & 5.01 & 28.12 & 4.37 & 0.17 \\
   \hline
 \end{tabular*}
\end{table}

\begin{figure}
	\begin{center}
    \hspace{-0.7cm}\includegraphics[scale=0.46]{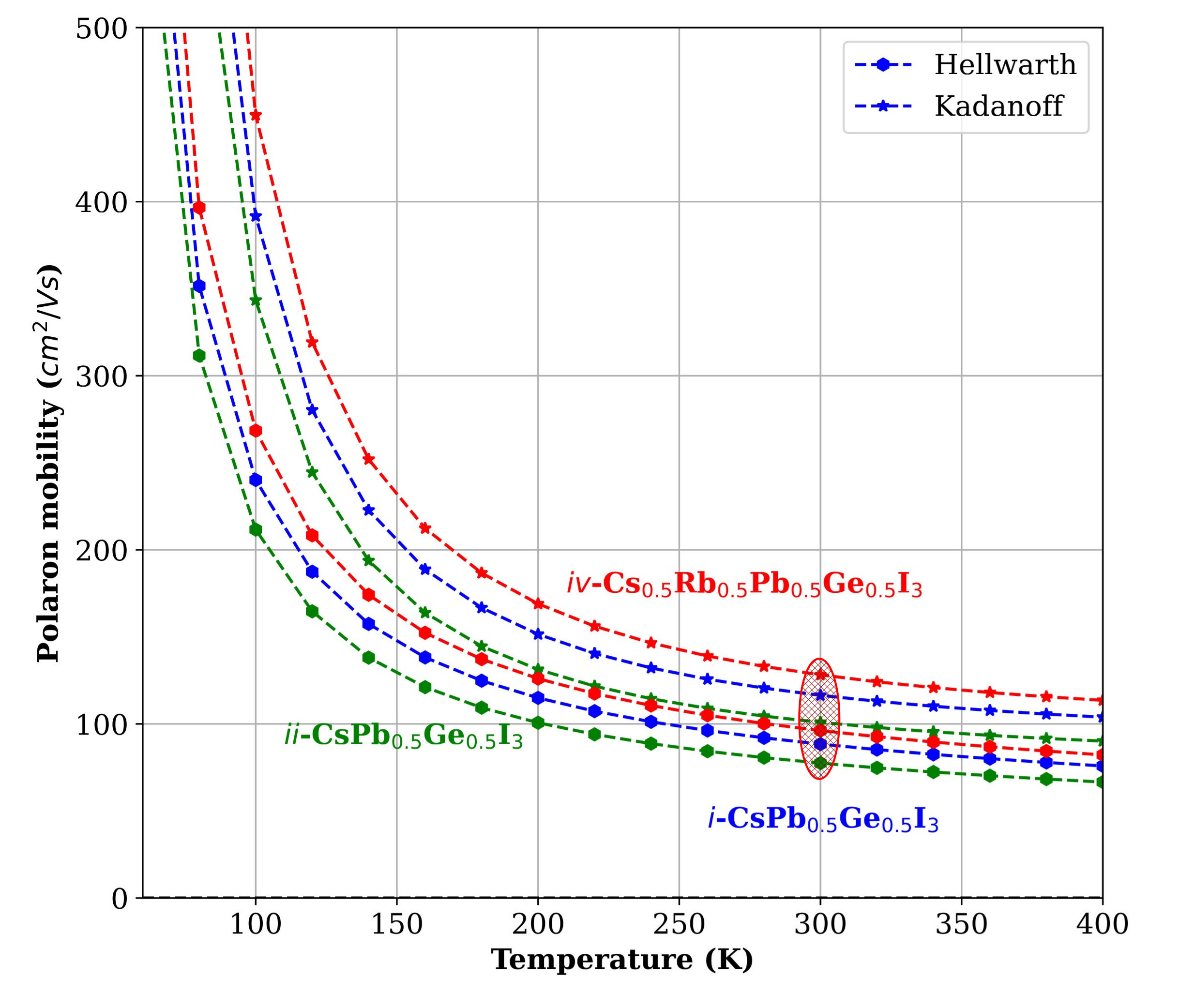}
    \caption{Polaron mobility as a function of temperature for configurations $i$, $ii$, and $iv$.}
    \label{pol_mob}
    \end{center}
\end{figure}

Using the parameters in Table \ref{feymanP_parameters}, an approximate estimate of temperature-dependent polaron mobility for 3 configurations has been determined based on the Feynman's variational model for Fr{\"o}hlich polaron 
\cite{feynman1955slow, osaka1959polaron, frost2017calculating, polaronf} and is presented in Figure \ref{pol_mob}. Mobility is calculated in two ways: Hellwarth model \cite{hellwarth1999mobility} (by contour integrating the self-energy of perturbed polaron) and Kadanoff's Boltzmann equation approximation \cite{kadanoff1963boltzmann}. The Hellwarth mobility is given by the expression
\begin{equation}
    \mu_{H} = \left(\frac{w}{v} \right)^{3} \frac{3e}{m_{b}^{*}} \frac{\sqrt{\pi}sinh(\beta/2)}{\omega_{eff}\alpha_{el-ph}\beta^{\frac{5}{2}}}K^{-1}
\end{equation}
The detailed expressions for $K$, $\beta$, $v$ and $w$ are given in reference \cite{frost2017calculating}. Similarly, the Kadanoff mobility is expressed as
\begin{equation}
    \mu_{K} = \left(\frac{w}{v} \right)^{3} \frac{e}{2m_{b}^{*}} \frac{exp(\beta)}{\omega_{eff}\alpha_{el-ph}} exp\left(\frac{v^2-w^2}{w^2v} \right)
\end{equation}
These models consider the band structure only in an effective mass approximation. From the expressions, it is clear that the electron-phonon coupling constant is inversely proportional to the polaron mobility. Large values of Fr{\"o}hlich coupling constant imply incoherent, low mobility ($<$ 1 cm$^{2}$/Vs) small polaron transport \cite{ghosh2020polarons}. The mobility at 300 K for 3 configurations is around 100 cm$^{2}$/Vs with the highest value for $iv$-Cs$_{0.5}$Rb$_{0.5}$Pb$_{0.5}$Ge$_{0.5}$I$_{3}$. Better results can be expected by using the effective phonon frequency from renormalized phonon frequencies. 
Furthermore, using the effective mass obtained at PBE+SOC level gives mobility values higher than the reported result of CsPbI$_{3}$. The mobilities decrease with an increase in temperature. Also, there is a small decrease in mobility with a reduction in symmetry: from $i$-higher symmetry cubic phase to $ii$-lower symmetry tetragonal phase. At the core of Kadanoff treatment is the calculation of the rate of emission and absorption of optical phonons. For the electron-polaron at a temperature of 300 K, the scattering time ($\tau$ = 1/$\Gamma$, where $\Gamma$ is the scattering rate) is 0.10, 0.10, and 0.11 ps for the configurations $i$, $ii$, and $iv$ respectively.

\begin{figure}
    \subfigure[]{\hspace{-0.1cm}\includegraphics[scale=0.42]{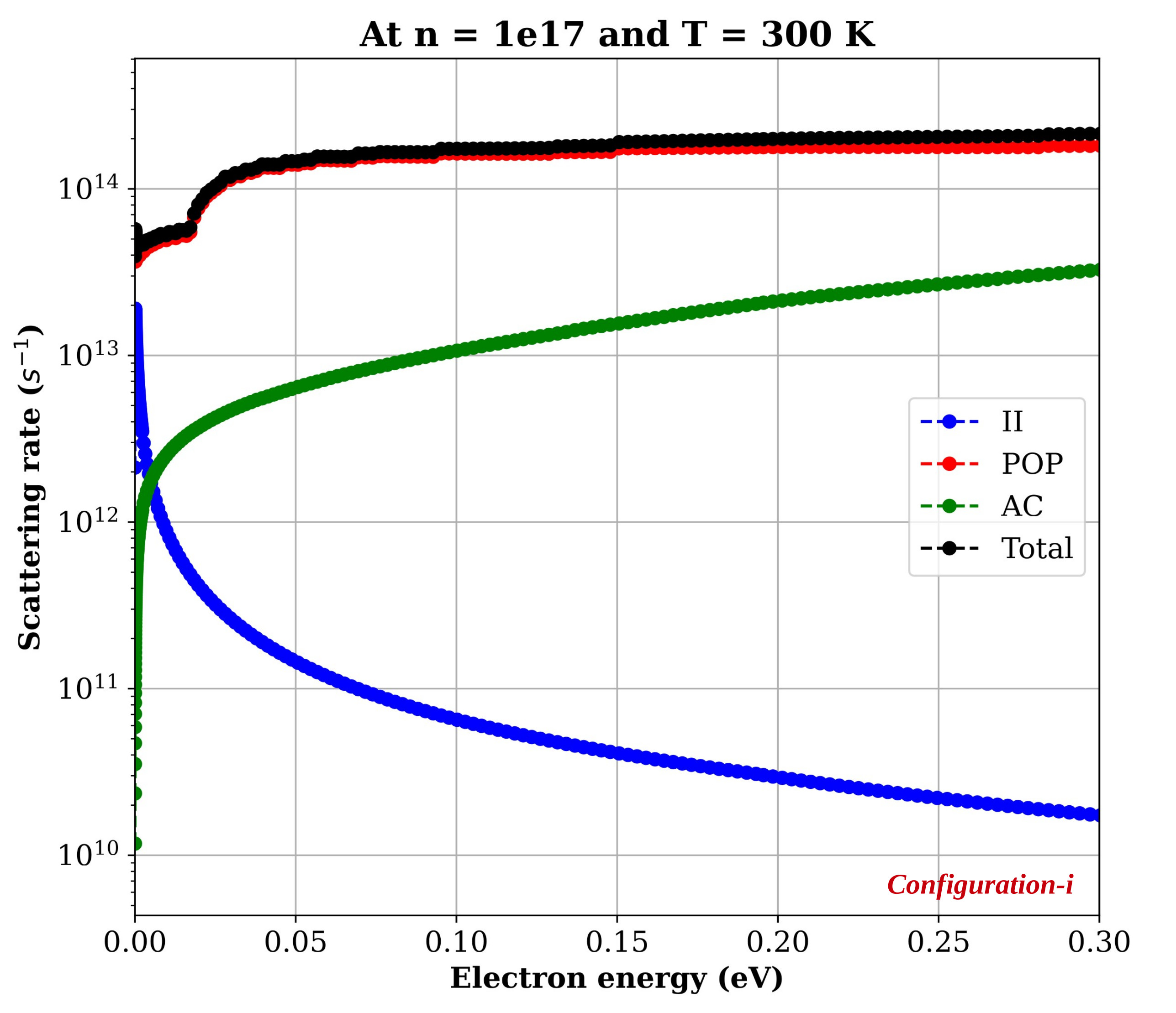}} \\
    \subfigure[]{\hspace{-0.2cm}\includegraphics[scale=0.40]{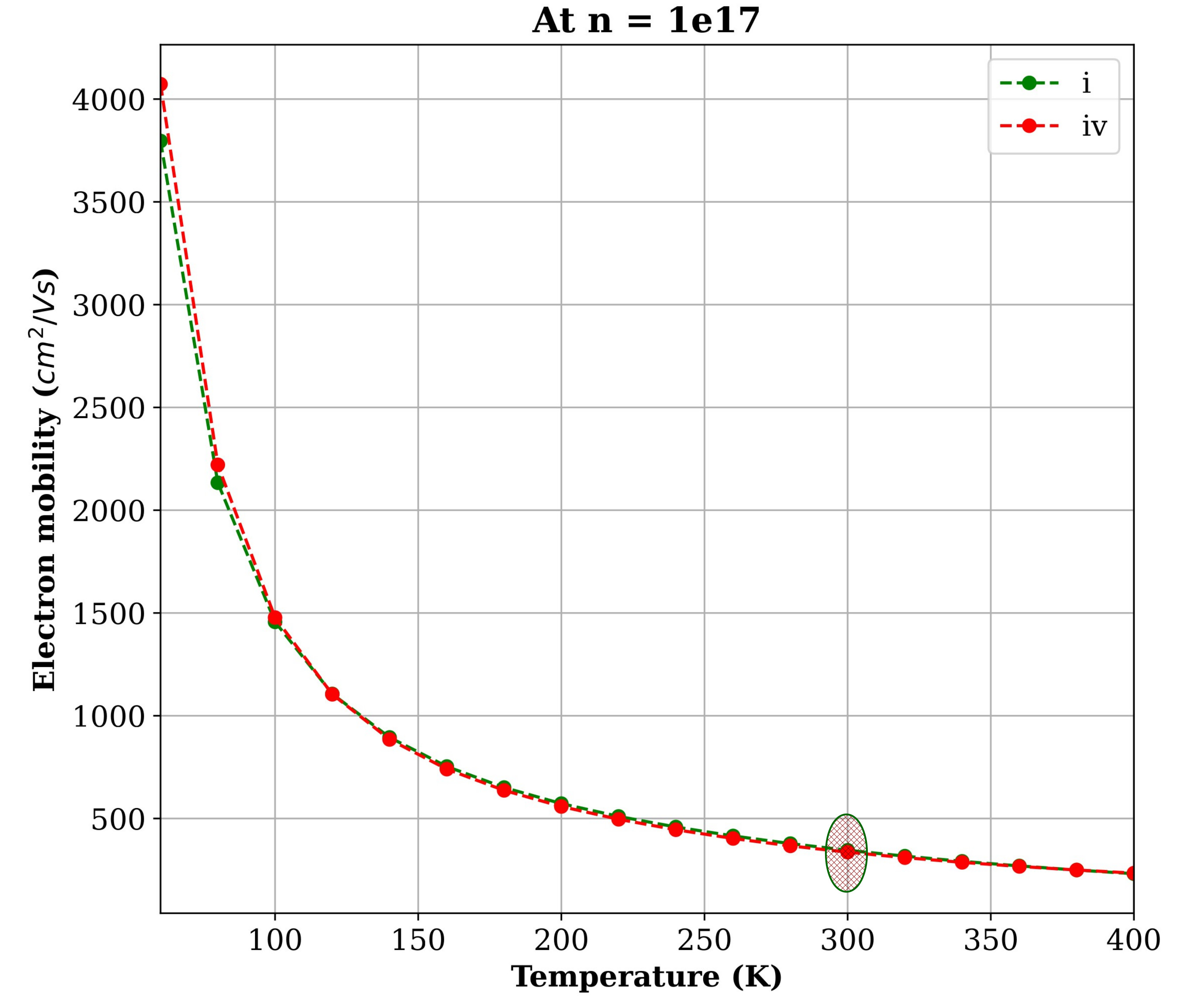}}
    \caption{(a) Total scattering rates of configuration $i$ with contributions of II, POP and AC scattering mechanisms. (b) Mobility of configurations $i$ and $iv$ as a function of temperature at a carrier concentration of 1$\times$10$^{17}$cm$^{-3}$.}
    \label{scattering_mobility}
\end{figure}
Since charge transport is crucial to the performance of solar cells, we made an alternate attempt to determine the mobility of configurations $i$ and $iv$ using Rode's iterative algorithm for solving the Boltzmann transport equation. But there are certain difficulties associated with calculation of input parameters for transport code (AMMCR), hence, these are approximate estimations. The primitive cell of configuration $i$ has 10 atoms and configuration $iv$ has 40 atoms. With increasing number of atoms, the mobility calculation becomes expensive. Another main hurdle is, that these configurations are not dynamically stable and have soft optical modes in their phonon dispersion at 0 K. This will cause errors in the determination of mobility. Earlier work by Su et al. \cite{su2021stability} to determine mobility of the parent CsPbI$_{3}$ in different phases ($\alpha$, $\beta$, and $\gamma$) using the electron-phonon coupling matrix element, skipped $\alpha$-phase due to the presence of multiple soft modes in the phonon dispersion spectrum. It is beyond the scope of this work to determine the temperature-dependent phonon dispersion. Here, mobility is evaluated at the frequency of the most polar mode and compared. We used the electronic and ionic contributions to the dielectric function at the PBE+U+SOC level in our mobility calculations. The high and low frequency dielectric constants of $\alpha$-CsPbI$_{3}$ reported by Su et al. are $\epsilon_{\infty}$ = 5.59 and $\epsilon_{s}$ = 29.77 \cite{su2021stability}. Our values are in a similar range. One difference from Frost's implementation \cite{frost2017calculating} of mobility determination is in the treatment of band structure. Here, the band structure is calculated for a highly dense k-mesh around the CBM and the average electron energies are expressed as a function of distance from the band edge extremum (CBM) \cite{mandia2021ammcr}. Then, analytical fitting of band near the conduction band edge is carried out with a sixth degree polynomial to obtain carrier group velocity, which is further used to calculate different scattering rates. Three scattering mechanisms are included in transport calculations, and they are polar optical phonon (POP), acoustic deformation potential (AC) and ionized impurity (II) scattering. The scattering rates for configuration $i$ are provided in Figure \ref{scattering_mobility}. It is seen that POP scattering is dominant in this mixed perovskite composition, and mobility is limited by them. 

The ab-initio input parameters for the calculation of mobility are provided in the supplemental material (Table S1). We observed that the mobility of configurations $i$ and $iv$ is close to each other at 300 K, which implies that Rb does not change the transport characteristics of CsPb$_{0.5}$Ge$_{0.5}$I$_{3}$, though it improves the stability (lower Gibb's free energy). At room temperature (RT), the local mobility of all-inorganic black $\gamma$-phase CsPbI$_{3}$ ($\gamma$-CsPbI$_{3}$) thin film is 270 $\pm$ 44 cm$^{2}$/Vs, as recorded using ultrafast terahertz spectroscopy \cite{zhang2021highly}. In the report by Frost, the polaron mobility (Hellwarth scheme) of $\alpha$-phase is calculated to be 258 cm$^{2}$/Vs at RT. These mobility values are much higher than those reported for hybrid organic-inorganic halide perovskites (MAPbI$_{3}$ and FAPbI$_{3}$) \cite{ponseca2014organometal, oga2014improved, piatkowski2016unraveling, rehman2015charge}. For both configurations ($i$ and $iv$), we obtained electron mobility (of 345 and 336 cm$^{2}$/Vs at RT and a carrier concentration of 1$\times$10$^{17}$ cm$^{-3}$) higher than the theoretically reported value of CsPbI$_{3}$. The low $\epsilon_{\infty}$ and high $\Delta\epsilon$ = $\epsilon_{s}$ - $\epsilon_{\infty}$ could be responsible for the higher values of mobility compared to CsPbI$_{3}$. In other words, it can be concluded that there is an improvement in stability without compromising the optoelectronic properties of CsPbI$_{3}$ with Ge-alloying and Rb-alloying at 50-50 compositions.

\section{Conclusions}
In summary, we investigated the effect of multi-site alloying on the thermodynamic stability and optoelectronic properties of inorganic halide perovskite, CsPbI$_{3}$, employing first principles calculations in combination with the cluster expansion method. The results show that, for both CsPb$_{0.5}$Ge$_{0.5}$I$_{3}$ and Cs$_{0.5}$Rb$_{0.5}$Pb$_{0.5}$Ge$_{0.5}$I$_{3}$, it is the configurational entropy that improves stability at finite temperatures. Their favorable direct band gaps with allowed optical transition from VBM to CBM (from transition probability) are an added advantage in solar cell applications. These mixed configurations are found to have large polarons which are envisioned to play a major role in charge extraction and recombination dynamics. To meet the growing demand for durable and stable perovskite-based optoelectronic materials and devices, alloying at multiple inequivalent sites (not just A and B, but A, B, and X or multiple elements at one site) is a tenable way to maximize the configurational entropy, thereby improving thermodynamic stability.
\section*{Acknowledgement}
We would like to thank the Korean National Supercomputing Center for providing computational resources through an R\&D Innovation Support Program (Grant No: KSC-2022-CRE-0510 \& KSC-2023-CRE-0342). NAK also would like to thank Dr. A. K. Mandia for useful discussions.
\bibliography{Ref1}
\end{document}